\documentclass[preprint,showpacs,preprintnumbers,amsmath,amssymb]{revtex4}

\usepackage{graphicx}
\usepackage{dcolumn}
\usepackage{bm}

\begin{document}

\preprint{submitted to Phys. Rev. E}

\title{Beam interactions in one-dimensional saturable waveguide arrays}

\author{Milutin Stepi\'c}
 \altaffiliation[Also at ]{Vin\v ca
Institute of Nuclear Sciences, P.O.B. 522, 11001 Belgrade,
Serbia.} \email{milutin.stepic@tu-clausthal.de}

\author{Eugene Smirnov}
\author{Christian E. R\"uter}
\author{Liv Pr\"onneke}
\author{Detlef Kip}%
 \affiliation{Institute of Physics and Physical Technologies, Clausthal University of Technology, 38678 Clausthal-Zellerfeld, Germany
}%

\author{Vladimir Shandarov}
\affiliation{State University of Control Systems and
Radioelectronics, 40 Lenin Ave., 634050 Tomsk, Russia}

\date{\today}

\begin{abstract}
The interaction between two parallel beams in one-dimensional
discrete saturable systems has been investigated using lithium
niobate nonlinear waveguide arrays. When the beams are separated
by one channel and in-phase it is possible to observe soliton
fusion at low power levels. This new result is confirmed
numerically. By increasing the power, soliton-like propagation of
weakly-coupled beams occurs. When the beams are out-of-phase the
most interesting result is the existence of oscillations which
resemble the recently discovered Tamm oscillations.
\end{abstract}

\maketitle

\section{Introduction}

There is a growing interest in routing, guiding, and manipulating
light by light itself. Such an all-optical concept can be
accomplished through the interaction of self-guided beams, which
are often called spatial solitons [1,2]. These localized
structures are found to exist in various settings such as, for
example, plasmas [3], Josephson junctions[4], molecular chains
[5], and in nonlinear optics [6]. In the latter case, the
refractive index profile induced by a soliton beam exactly
balances the inherent beam divergence due to diffraction. It has
been demonstrated that logic gates and all-optical switching are
possible exploiting the interaction of a couple of parallel beams
[7-9]. Here the mutual interaction of two beams, resulting from
the additional contribution to the induced refractive index change
of the overlapping input fields, depends crucially on their
relative phase. When the beams are in-phase they attract each
other while repulsion occurs if they are $\pi$ out of phase
[10-12]. In intermediate cases there appears an energy transfer
between the two beams [9,13].

Homogeneous nonlinear waveguide arrays (NWA) represent a periodic
arrangement of parallel, weakly coupled waveguides. They have been
realized in semiconductors [14], photorefractive crystals [15-17],
and nematic liquid crystals [18], to mention a few. NWA could be,
for example, used for switching [19], passive mode locking [20],
and tapered laser arrays [21]. Interactions of two initially
parallel beams have been, up to date, investigated only in AlGaAs
waveguide arrays exhibiting a cubic, self-focusing Kerr-like
nonlinearity [22,23]. In this work, we focus on interactions of
parallel beams in one-dimensional NWA in lithium niobate. It is
well-known that this photovoltaic photorefractive material
exhibits a self-defocusing nonlinear response that has a saturable
nature [24]. Our main findings are the demonstration of fusion of
two in-phase beams and oscillations in the case of out-of-phase
beams. We demonstrate numerically that these effects exist in
cubic self-defocusing discrete media, too.

The paper is organized as follows. In Sec.~II we describe our
experimental setup. Sec.~III is devoted to the interactions of
in-phase beams. Here, we present our experimental results which
have been confirmed numerically by simulations based on a
nonlinear beam propagation method. For the sake of completeness,
we add the corresponding numerical results for cubic
self-defocusing media. In Sec.~IV we analyze numerically the
interactions of out-of-phase beams in both saturable and cubic NWA
while the conclusions are given in Sec.~V.

\section{Experimental methods}

Our sample is a 27 mm long, x-cut lithium niobate crystal.
Permanent channel waveguides are fabricated by Ti indiffusion. A
lithographically patterned Ti layer with a thickness of 10\,nm is
annealed for 2 hours at a temperature $T=1040\,^\circ$C. The
sample is thereafter surface doped by Fe (5.6 nm Fe layer,
annealed for 24 hours at $T=1060\,^\circ$C). This additional
doping serves to enhance the photorefractive effect. Each channel
is 4\,$\mu$m wide and forms a single-mode waveguide for TE
polarized green light. The distance between adjacent channels is
4.4\,$\mu$m, which results in a lattice period of
$\Lambda=8.4\,\mu$m. The input and output facet of the sample are
finally polished to optical quality.

Our experimental setup is sketched in Fig.~1. The light source is
a Nd:YVO$_4$ laser that provides single-frequency output at a
wavelength $\lambda$ = 532\,nm. We form a 3\,cm wide quasi-plane
wave by means of a beam expander ($20\times$ microscope lens and
collimation by a second lens with focal length $f=200\,$mm). To
excite different light patterns on the input face of the sample,
an adequate amplitude mask (titanium on glass substrate covered by
photo resist) has been fabricated using a laser beam writer. The
mask is placed in front of a $40\times$ microscope lens that
images two illuminated holes of the mask with a diameter of
$2r=142\,\mu$m separated by $d=704\,\mu$m. This mask transmits two
in-phase beams which are adjusted by virtue of the microscope lens
in such a way to excite only two channels of the array. As the
coupling in our NWA is relatively weak we restrict our study to
the case in which these two channels are separated by one channel.
Green light from the output facet is collected by another $20
\times$ microscope lens and imaged onto a CCD camera.

\begin{figure}
\includegraphics[width=8.2cm]{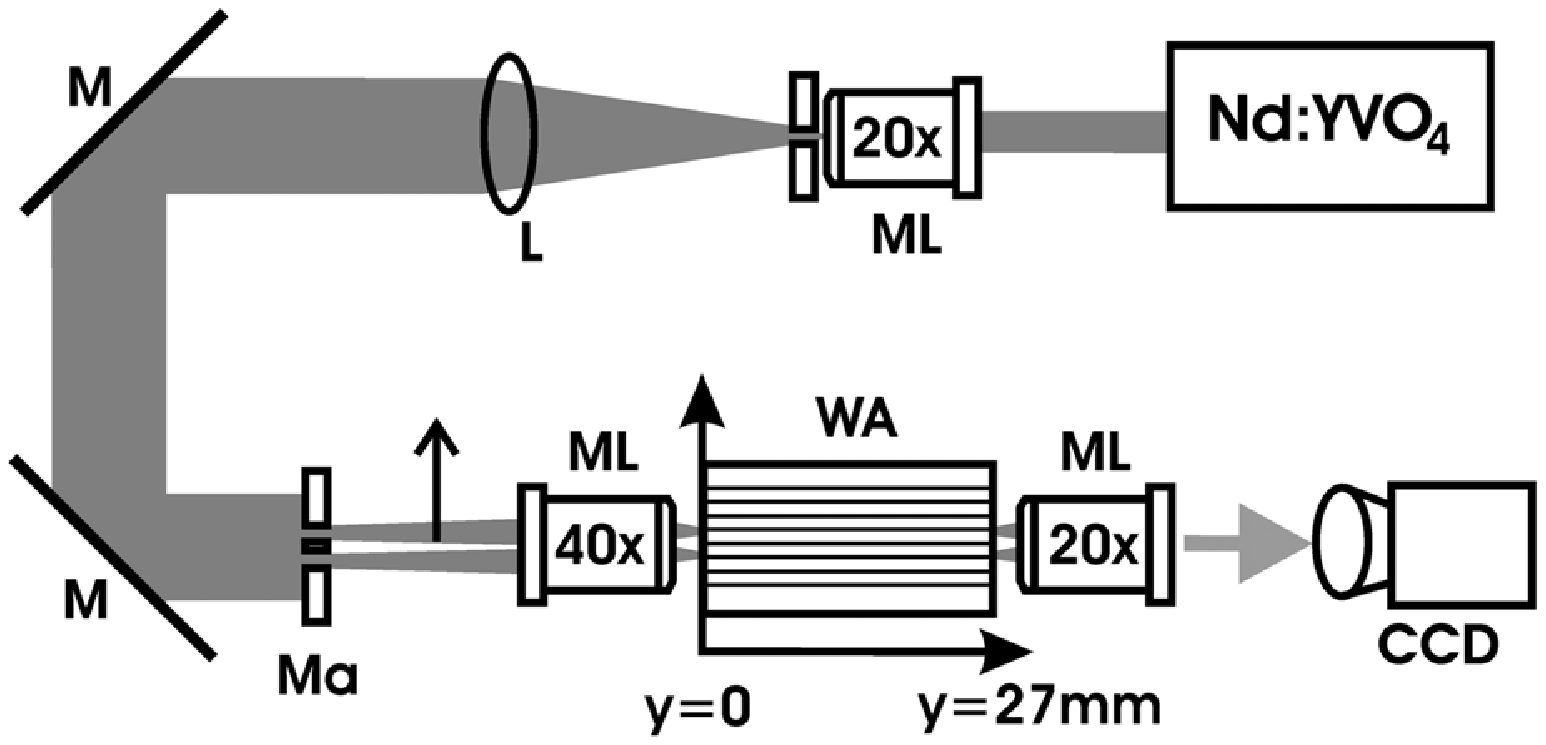}
\caption{\label{Fig1}Experimental setup: ML, microscope lens; L,
lens; M, dielectric mirrors; Ma, mask; WA, waveguide array; CCD,
CCD camera.}
\end{figure}

\section{Interactions of in-phase beams}

As is well established, scalar wave propagation in a nonlinear
one-dimensional WA can be modelled within a paraxial approximation
by:
\begin{equation}
i\frac{\partial E}{\partial y}+ \frac{1}{2k}\frac{\partial^2
E}{\partial z^2}+k\frac{n(z)+\Delta n_{nl}}{n_s}E=0~.
\end{equation}
The propagation coordinate is along the $y$-axis, the amplitude of
the electrical field is denoted by $E$, while $k=2\pi n_s/\lambda$
represents the wave number. Here, $\lambda$ is the wavelength of
the used light in vacuum while $n_s=2.2341$ is the extraordinary
refractive index of our lithium niobate substrate. The
periodically modulated refractive index which defines the
nonlinear WA is denoted by $n(z)$ while $\Delta n_{nl}$ is the
nonlinear refractive index change ($\Delta n_{nl}<<n_s$). The
periodically modulated refractive index can be well approximated
by $n(z)=2.2341+0.01035\cos^2(\pi z/\Lambda)$.

In the following we investigate the interaction of two in-phase
beams separated by one channel on the input facet. In Fig.~2 we
give typical examples of experimentally observed discrete
diffraction and nonlinear interaction of the two beams,
respectively. Here each of the beams has an optical power
$P\approx 7\,\mu$W. An image of discrete diffraction of two beams
from the output facet of the lithium niobate NWA is presented in
Fig.~2a. In parts b) and c) the corresponding images of discrete
diffraction when one of the beams is blocked are presented. These
light distributions serve us to ensure the correct input
excitation with straight propagation within the array (zero
transverse wave vector component). In Fig.~2d we monitor the
temporal evolution of the interaction of the two beams. After an
initial stage of discrete diffraction and a short transient regime
a stable, steady state, two-hump structure is formed within a few
minutes.

\begin{figure}
\includegraphics[width=8.2cm]{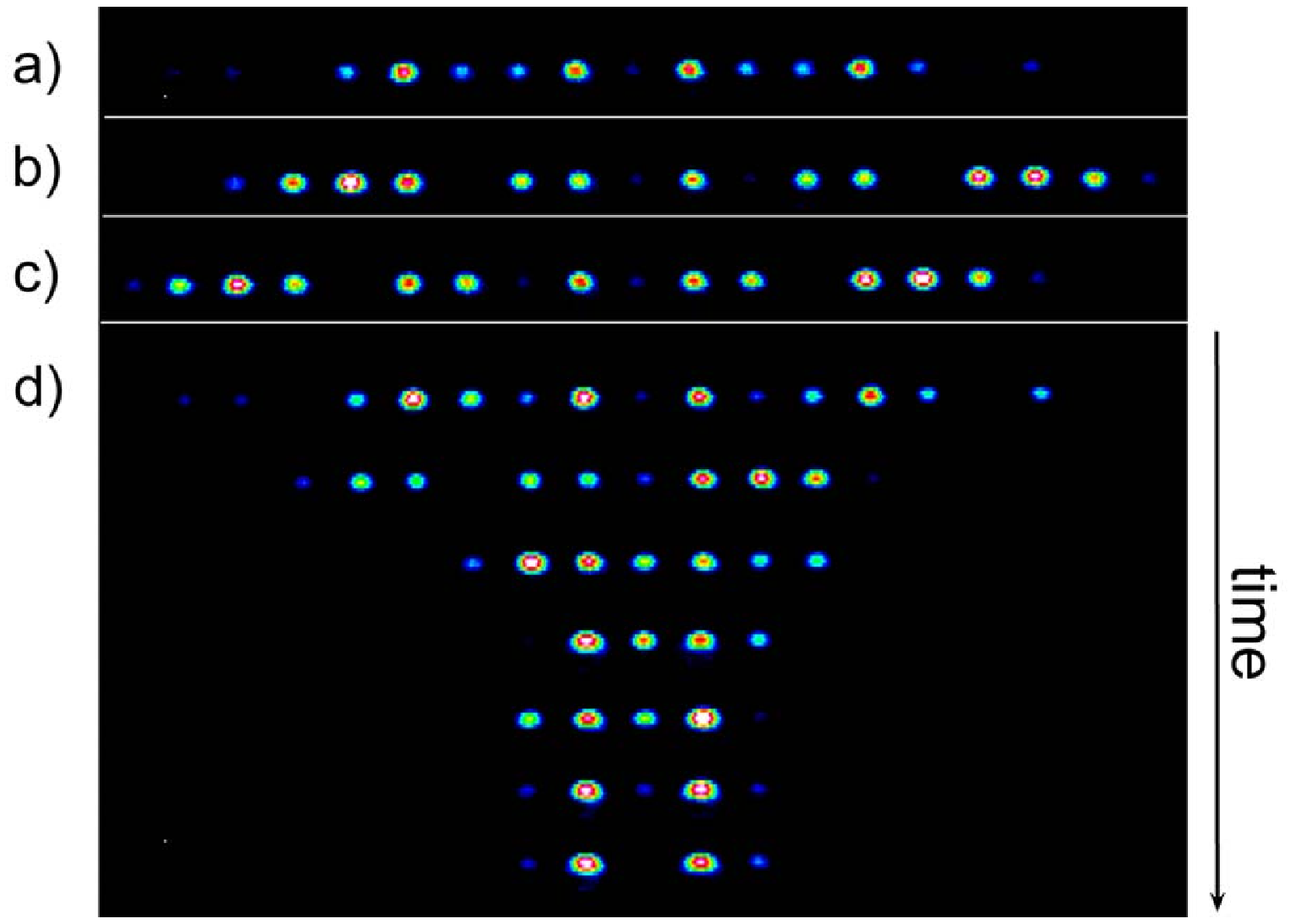}
\caption{\label{Fig2} (Color online) Discrete diffraction (a-c)
and nonlinear interaction (d) of two co-propagating in-phase
beams. The photographs show the corresponding images on the output
facet of the NWA when: a) both beams are present; b, c) one beam
is blocked, and d) time evolution of two-beam interaction for an
input power of $P\approx 7\,\mu$W.}
\end{figure}

In Fig.~3 we present experimental results for different power
levels of the two parallel beams and compare the obtained results
with numerical modelling. For this we solve Eq.~(1) numerically by
using a nonlinear beam propagation method (BPM). We used the
parameters of our WA and a saturable defocusing nonlinearity of
the form
\begin{equation}
\Delta n_{nl}=\Delta n_0 I/(I+I_d)~,
\end{equation}
with amplitude $|\Delta n_0|= 3\times 10^{-4}$ and an intensity
ratio $r=I/I_d$, where $I_d$ is the so-called dark irradiance and
$I$ is the light peak intensity.

In Fig.~3 a) linear discrete diffraction is measured and compared
with theory, yielding the corresponding coupling constant of
$L_c=3.4\,$mm of our sample. Experimentally, in the low power
regime ($P\approx 0.5\,\mu$W) in b) we observe soliton fusion in
the central channel of the array, in good agreement with the BPM
results. This process is absent in the cubic case [23]. On the
other hand, soliton fusion has been observed in both bulk and
planar waveguide photorefractive crystals exhibiting a saturable
nonlinearity [25,26]. The formed structure possess a highly
symmetric form of strongly localized mode A [27,28]. In the regime
of mediate power in c) it is possible to obtain almost
independent, soliton-like propagation of the two beams, as
observed for single-channel excitation [28]. Here one can observe
weak oscillations which result in light localization either in the
central element or in its first neighbors, which can be understood
by the remaining weak evanescent coupling of the two parallel
waveguides. This oscillatory behavior has been reported in
Ref.~23, too. For higher power and thus a stronger effect of
saturation we observe a widening of the formed structure
(Fig.~3d), again in good agreement with numerics.

\begin{figure}
\includegraphics[width=8.2cm]{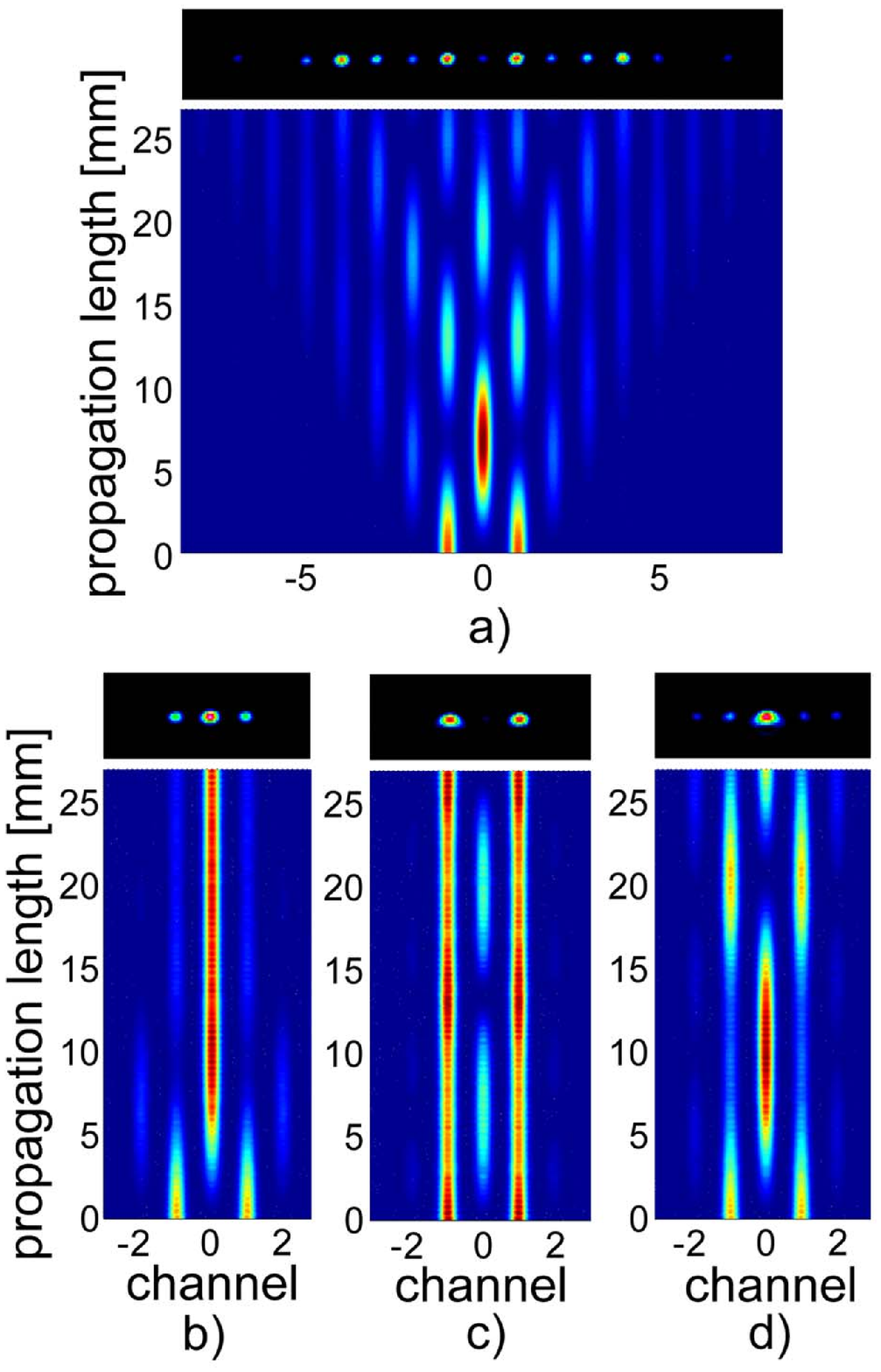}
\caption{\label{Fig3} (Color online) (Color online) Comparison of
in-phase interaction for different input powers and saturable
nonlinearity. a) Discrete diffraction. b) Soliton fusion:
$P\approx 0.5\,\mu$W, $r=0.43$. c) Soliton-like propagation:
$P\approx 7\,\mu$W, $r=4.62$. d) A wide structure: $P\approx
25\,\mu$W, $r=30$.}
\end{figure}

It is well known that photorefractive crystals such as strontium
barium niobate and lithium niobate have a non-instantaneous
nonlinear response [29]. Depending on light intensities, build-up
times in these materials range from a few milliseconds to a few
minutes or even hours. Thus, we are able to perform a specific
read-out of the light induced structures that have been presented
in Figs.~2 and 3. For this, after recording of stationary
refractive index changes we block one of two writing beams. The
residual beam is still able to "see" the former light-induced
refractive index change, as demonstrated in Fig.~4: The two
induced waveguides are evanescently coupled to each other, leading
to partial energy transfer from one channel to the other.

\begin{figure}
\includegraphics[width=8.2cm]{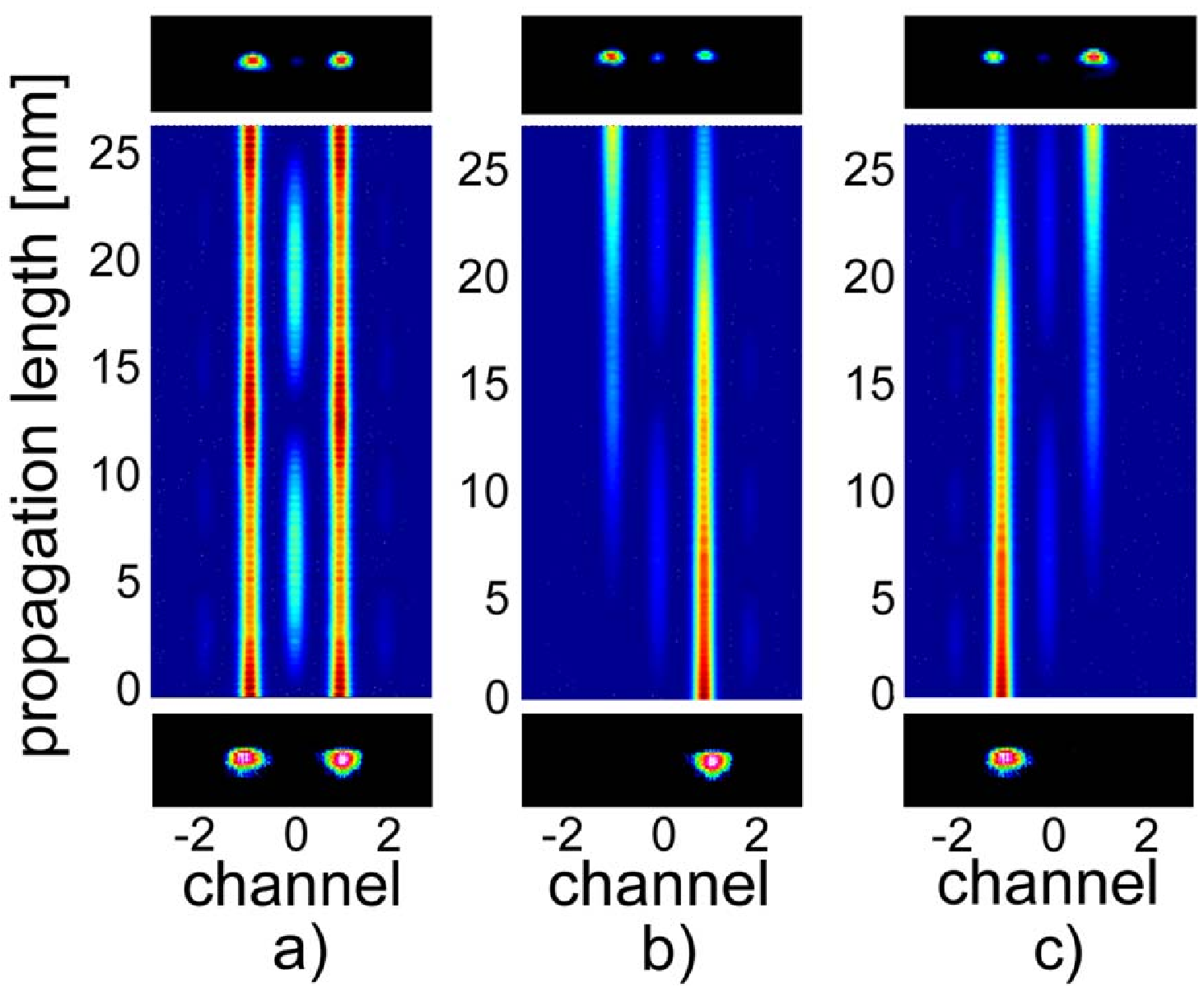}
\caption{\label{Fig4} (Color online) Read-out of light-induced
structures. Upper parts: images from the input facet (top) and the
corresponding images from the output facet (bottom). Lower parts:
numerical simulation of build-up (a) and single beam propagation
(b,c) in the induced structure. a) Initial two-beam interaction;
b) and c) read-out of the induced structure with a probe beam
coupled in the left or right input channel, respectively.}
\end{figure}

The fact that the output from the array can be controlled by
changing only the power of two beams may be attractive for fast
all-optical gating. Therefore, we perform simulations in WA with
instantaneous cubic nonlinearity [2,11,22,23]. In this case we
have $\Delta n_{nl}=\Delta n_0 I$. We arbitrarily take, as before,
$|\Delta n_0|=3\times 10^{-4}$ and the above mentioned data for
our waveguide array. Most important results shown in Fig.~5 [the
fusion of solitons in a) and the soliton-like propagation of two
beams in b)] reveal that the interaction of two parallel in-phase
beams is independent of the type of nonlinearity.

\begin{figure}
\includegraphics[width=5.5cm]{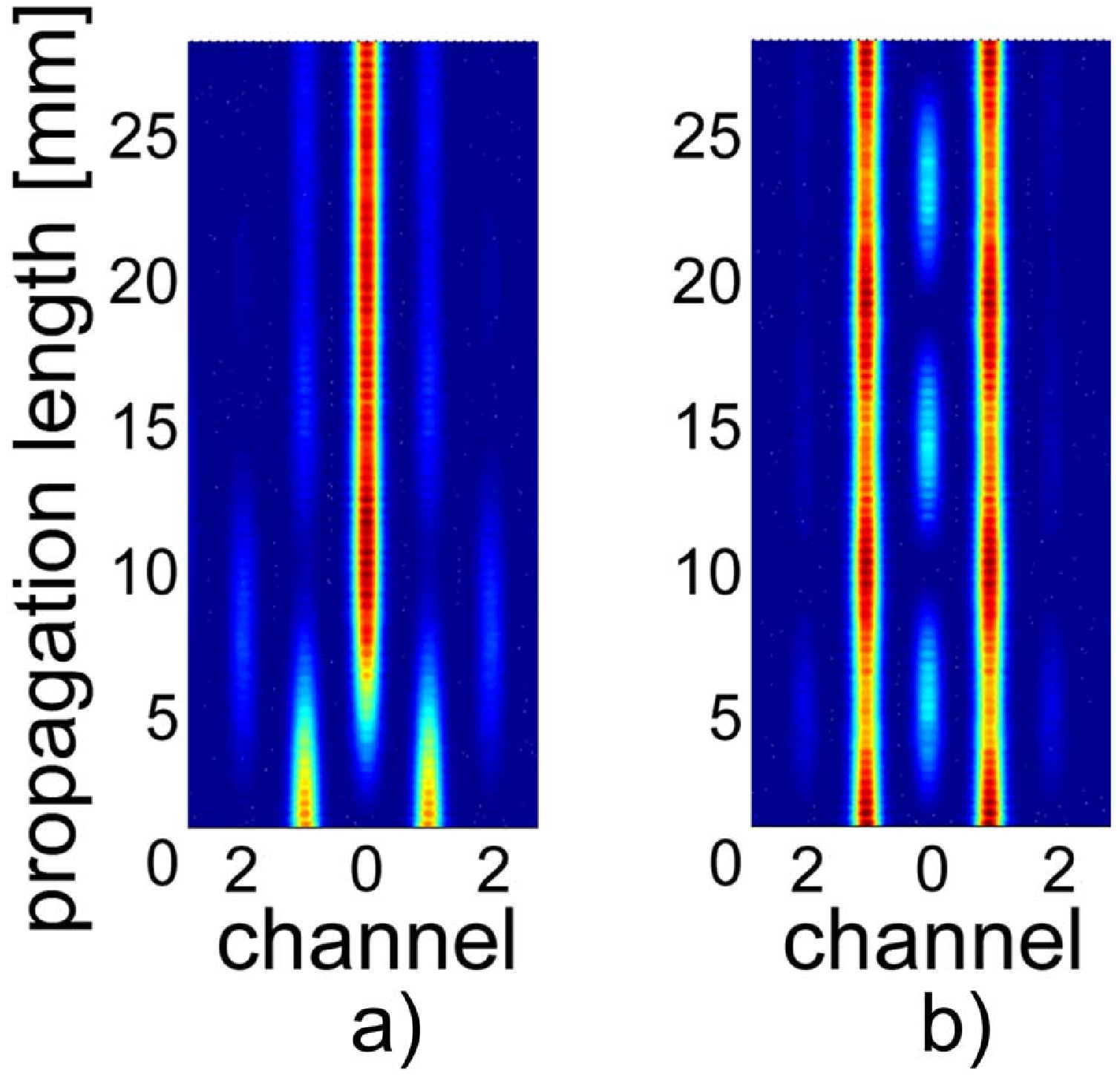}
\caption{\label{Fig5} (Color online) Interaction of in-phase beams
for cubic nonlinearity: a) Fusion of two beams with $\Delta
n_{nl}=2.8\times 10^{-5}$. b) Nearly independent propagation of
two soliton-like beams for $\Delta n_{nl}=6.2\times 10^{-5}$.}
\end{figure}

\section{Interactions of out-of-phase beams}

In this Section we study numerically interactions of two
out-of-phase beams in a WA for both saturable and cubic defocusing
nonlinearities. From the investigation of bulk and waveguide
arrays with cubic nonlinearity, it is known that out-of-phase
parallel beams will repel each other [10-12,23].

Some examples of our numerical results are shown in Fig.~6. The
first four pictures are devoted to the saturable case while the
last two are for a cubic (Kerr) nonlinearity. In the low power
regime (a) one can recognize the expected repulsive behavior of
this interaction while in the high power regime (d, f) we find a
practically independent propagation of two soliton beams. These
results are in full agreement with the corresponding findings from
self-focusing discrete media with cubic nonlinearity [23].
However, the oscillations presented in (b, c, e) that are observed
for mediate power levels are a completely new phenomenon which
resembles the recently observed Tamm oscillations at the interface
between a homogeneous substrate and a WA [30]. An intuitive
explanation is that out-of-phase beams are reflected back from the
specific channel for which the Bragg condition is fulfilled. Bloch
oscillations [31] occur as a special case of Tamm oscillations
when the repulsive potential is a linear function of the distance
from the edge of the array. These oscillations have a promising
role in all-optical switching at low power level as reported, for
example, in Ref.~32. Here it is important to mention that in this
Section we use the corresponding data of another sample which has
an approximately three times shorter coupling length of
$L_c=1.1$\,mm [33]. Namely, as has been shown in Ref.~30, the
period of Tamm oscillations increases with the growth of the
coupling length. Thus, our 27\,mm long iron-doped sample is still
too short to observe a clear oscillatory behavior. Also, the
corresponding data for $\Delta n_{nl}$ used for the saturable case
are around the maximally achievable value of nonlinear refractive
index changes in lithium niobate, which is of the order of
$1\times 10^{-3}$. For lower values of $\Delta n_{nl}$ it remains
difficult to observe both multiple oscillations and soliton-like
propagation of two beams.

\begin{figure}
\vspace*{6mm}
\includegraphics[width=8.2cm]{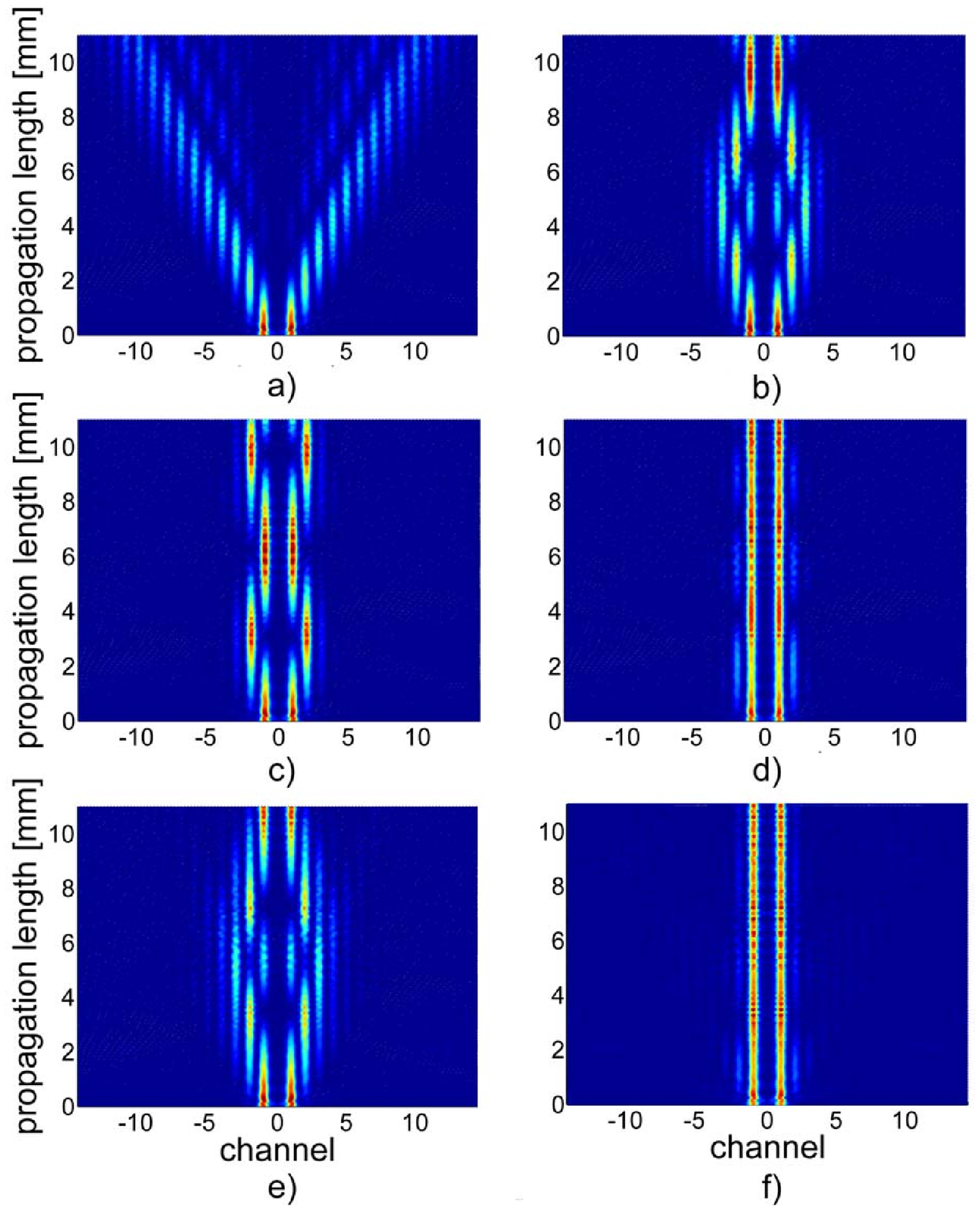}
\caption{\label{Fig6} (Color online) Interaction of two
out-of-phase beams in saturable self-defocusing media (a-d) and
cubic self-defocusing media (e,f): a) repulsion for $\Delta
n_{nl}=9\times 10^{-4}$ and $r=0.6$; b) one oscillation for
$\Delta n_{nl}=9\times 10^{-4}$ and $r=2.61$; c) two oscillations
for $\Delta n_{nl}=9\times 10^{-4}$ and $r=5$; d) undisturbed
propagation of two soliton-like beams for $\Delta
n_{nl}=1.15\times 10^{-3}$ and $r=5$; e) one oscillation for
$\Delta n_{nl}=3.65\times 10^{-4}$; and f) two soliton-like beams
for $\Delta n_{nl}=5\times 10^{-4}$.}
\end{figure}

\section{Conclusion}

The interaction between two parallel beams in one-dimensional
nonlinear waveguide arrays is investigated both experimentally and
numerically. As our iron-doped lithium niobate sample has rather
low coupling constant we concentrate on the case in which these
beams are separated by a single channel. We observe a complete
fusion of two in-phase beams at low power level. For higher input
power the interaction decreases and nearly independent propagation
of two separated solitons is observed. Another phenomenon which
does not have an analog in the self-focusing domain is the
oscillatory behavior of two out-of-phase beams. Both effects are
also obtained numerically in waveguide arrays exhibiting an
instantaneous cubic nonlinearity. Finally, our findings are of
considerable interest for all-optical gating and switching using
discrete soliton interaction in waveguide arrays.

\begin{acknowledgments}
This work has been supported by the German Federal Ministry of
Education and Research (BMBF, grant DIP-E6.1) and the German
Research Foundation (DFG, grant KI482/8-1)).
\end{acknowledgments}

\bibliography{apssamp}
{}
\end{document}